\documentclass[twocolumn, 10pt, notitlepage, twoside]{article}

\usepackage[utf8]{inputenc}
\usepackage[english]{babel}
\usepackage{csquotes}

\usepackage{geometry}
\usepackage{graphicx}

\usepackage{amsmath}
\usepackage[T1]{fontenc}

\usepackage[most]{tcolorbox}

\NewTColorBox[auto counter]{box1}{ s O{!htbp} }{%
  floatplacement={#2},
  IfBooleanTF={#1}{float*,width=\textwidth}{float},
  colframe=black!75!white,colback=black!3!white,title=Box 1: Cognitive tasks of untethered object perception,label=box:tasks,left=0pt,right=0pt,top=0pt,bottom=0pt
  }
 
\NewTColorBox[auto counter, use counter from=box1]{box2}{s O{t}}{%
  floatplacement={#2},
  IfBooleanTF={#1}{float*,width=\textwidth}{float},
  colframe=black!75!white,colback=black!3!white,title=Box 2: The binding problem,label=box:binding
  }

\usepackage[mathlines, switch]{lineno}
\modulolinenumbers[5]

\usepackage[allcolors=black]{hyperref}

\usepackage[autocite = superscript,backend=biber,style=nature, doi=false, url=false,isbn=false]{biblatex}
\AtBeginBibliography{\footnotesize}

\bibliography{bibliography.bib}

\usepackage{authblk}
\author[1,*]{Benjamin Peters}
\author[1,2,3,4,*]{Nikolaus Kriegeskorte}
\affil[1]{Mortimer B. Zuckerman Mind Brain Behavior Institute, Columbia University, New York}
\affil[2]{Department of Psychology, Columbia University, New York}
\affil[3]{Department of Neuroscience, Columbia University, New York}
\affil[4]{Department of Electrical Engineering, Columbia University, New York}
\date{}

\newcommand\blfootnote[1]{%
  \begingroup
  \renewcommand\thefootnote{}\footnote{#1}%
  \addtocounter{footnote}{-1}%
  \endgroup
}

\usepackage{fancyhdr}
\newcommand\shorttitle{Capturing the objects of vision with neural networks}
\newcommand\authors{Peters \& Kriegeskorte}
\usepackage{cuted}

\usepackage{caption}
\begin{document}
\title{\huge Capturing the objects of vision with neural networks}%
\maketitle

\begin{abstract}
Human visual perception carves a scene at its physical joints, decomposing the world into objects, which are selectively attended, tracked, and predicted as we engage our surroundings. Object representations emancipate perception from the sensory input, enabling us to keep in mind that which is out of sight and to use perceptual content as a basis for action and symbolic cognition. Human behavioral studies have documented how object representations emerge through grouping, amodal completion, proto-objects, and object files. Deep neural network (DNN) models of visual object recognition, by contrast, remain largely tethered to the sensory input, despite achieving human-level performance at labeling objects. Here, we review related work in both fields and examine how these fields can help each other. The cognitive literature provides a starting point for the development of new experimental tasks that reveal mechanisms of human object perception and serve as benchmarks driving development of deep neural network models that will put the object into object recognition.
\end{abstract}

\maketitle

\blfootnote{$^*$ correspondence should be addressed to \mbox{benjamin.peters@posteo.de} or \mbox{n.kriegeskorte@columbia.edu}}

Vision gives us a rapid sense of our surroundings that exceeds the information in the retinal image and provides a structured understanding of the scene. The structure imposed on the basis of prior knowledge is central to perception as an inference process \autocite{von_helmholtz_handbuch_1867, yuille_vision_2006} and to a causal and compositional understanding that enables us to consider counterfactuals and act intelligently \autocite{pearl_causality_2009}. The basic building blocks of our perceptual representation are objects. Our percepts include parts of objects that are occluded by other objects or behind us. Out of sight, for a mature primate, is not out of mind \autocite{piaget_construction_1954}. Relevant objects that become invisible remain represented, a memory trace, and may even be animated in our minds according to a rough approximation of the laws they obey in the world. 

Human behavioral researchers have quantitatively investigated these phenomena using a wide range of ingenious experimental paradigms. They have condensed the insights gained from the data in cognitive theories, which describe separate mechanisms for seeing stuff \autocite{adelson_seeing_2001} and seeing things \autocite{clowes_seeing_1971}. ``Stuff'' has come to refer to parts of the visual scene represented in terms of summary statistics \autocite{julesz_experiments_1975, simoncelli_natural_2001, rosenholtz_measuring_2007} that capture textures, materials, and perhaps categories at an aggregate level. ``Things'' are the objects that our brains pick out for individuated representation. An object representation may explicitly bind together the parts of each object and the image features each part accounts for \autocite{hoffman_parts_1984}. An object’s missing information may be filled in by inference using prior information \autocite{michotte_les_1964}. Cognitive scientists have described how bottom-up and top-down processes interactively determine the formation of a limited number of object representations that are accessible to higher cognition \autocite{rensink_dynamic_2000}.

The object representations may have a life of their own, simulating trajectories and interactions among objects to predict the future. Short of foreseeing the future, even being on time in representing the present requires prediction: to compensate for signalling delays in the nervous system. The perceived world emerges from the confluence in the inference process of prior information and present sensory signals \autocite{gregory_perceptions_1980, rock_indirect_1997}. Our brains combine past experience over multiple time scales to best predict the present and the future \autocite{clark_whatever_2013, friston_theory_2005, von_helmholtz_handbuch_1867, yuille_vision_2006}.

\begin{figure*}[ht]
\centering
\includegraphics[width=0.98\textwidth]{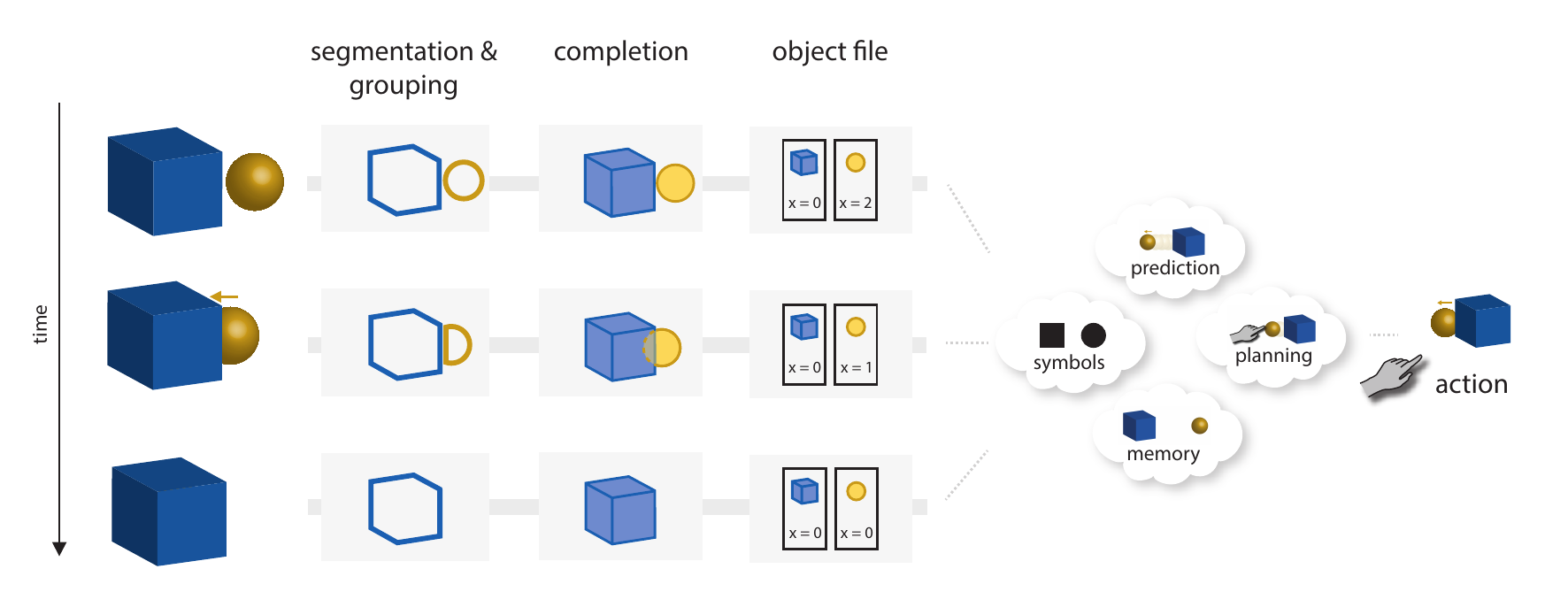}
\caption{\label{fig:overview} \textbf{Stages of untethering human visual object perception from the sensorium}. As the golden ball moves behind the blue box (left column from top to bottom), it is first unoccluded, then partially occluded, and finally fully occluded. It remains represented at the level of its object file even when fully invisible. The initial segmentation parses the scene into groups of features, each corresponding to one of the objects. Amodal completion may occur for partially occluded objects, completing the invisible portion of the object on the basis of short-term or long-term memory of its shape. A subset of the objects may be encoded in a non-retinotopic object-based representation (e.g., object-files). Object files can sustain information about the presence and properties of objects across temporary occlusions, untethering the object representations from the sensorium. Untethered object representations can be considered an interface between perception and symbolic thought, prediction, mental planning, and action.}
\end{figure*}

Cognitive scientists want to understand these dynamic and constructive inferences and the representations of objects in the human mind. Object representations abstract from the sensory features and cast the world as a composition of entities that can be acted on and named. This places object representations at the nexus of perception, action, and symbolic cognition (Fig. \ref{fig:overview}).

Engineers may not be interested in modeling the human mind. However, engineering, too, benefits from models that have concepts of objects, because they promise, for example, to enable a robot to understand the structure of the world, and to reason, plan, and act on this basis. For humans and machines alike, decomposing the world into objects may facilitate the modular reuse of learned knowledge and simplify complex inferences. An object-based representation provides a radical abstraction from the stream of sensory signals, a predictable scaffold of reality, and a basis for causal understanding. Building models with object-based representations is therefore a crucial challenge for engineering \autocite{van_steenkiste_perspective_2019, greff_binding_2020} as well as for cognitive science.

Parsing the world into objects requires an operational definition: What is an object? A key criterion is physical cohesion \autocite{spelke_principles_1990}. As \autocite{scholl_object_2007} put it: "If you want to know what an object is, just 'grab some and pull'; the stuff that comes with your hand is the object." This operational definition grounds objects in the physical structure of the world. Sensorimotor interactions, such as grabbing and pulling, may help us acquire the perceptual ability to parse the world into objects in early development \autocite{piaget_construction_1954}. They also continue to serve us in maturity, enabling us to confirm, through direct experiment, our perception that something is an object. The operational, "what if" nature of this definition reveals that objects are rooted in a causal understanding of physical reality \autocite{pearl_causality_2009}.

Object-based representations carve the scene at its physical joints. Reducing a million retinal signals to a few behaviorally relevant objects requires prior knowledge of the physical world, prior percepts from the present scene, and selection of what is relevant in light of the current behavioral goals. The present sensory evidence, then, does not solely determine the percept; it is just one of a number of constraints. Object representations, thus, \textit{untether} and emancipate perception from the stream of sensory signals.

Engineering has made substantial inroads toward this type of dynamic and constructive perceptual inference. The integration of sensory data over multiple timescales is captured by the Bayes filter, a recurrent mechanism that stores a compressed representation of recent experience for optimal representation of the present moment \autocite{sarkka_bayesian_2013}. Recurrent neural networks (RNNs) provide a universal model class for such inferences that can implement Bayes filters \autocite{deneve_optimal_2007}. However, getting RNNs to perform this kind of inference for natural dynamic vision (video) remains challenging. Computer vision therefore heavily relies on feedforward convolutional neural network models, which analyze each frame separately through a hierarchy of nonlinear transformations \autocite{lecun_backpropagation_1989, fukushima_neocognitron:_1980}. Feedforward deep convolutional neural networks can learn static mappings from images to category labels or structural descriptions of the scene. However, the representations in these models remain tethered to the input and lack any concept of an object. They represent things as stuff \autocite{geirhos_imagenet-trained_2019}. They cannot combine information over time so as to condition current perceptual inferences on past observations. They may also not be ideal for parsing scenes into objects. These limitations may explain why the performance of feedforward convolutional networks is somewhat brittle, breaking down when the models must generalize across domains \autocite{kansky_schema_2017}. The models lack what humans have: a generative structural and causal understanding of the world, to stabilize their perception \autocite{lake_building_2017, yildirim_integrative_2019, golan_controversial_2020}.

A generative mental model is a model of the process that generates the sensory data. A mind that employs a generative model is challenged to comprehensively explain all aspects of the sensory data, rather than taking a shortcut and selectively extracting only behaviorally relevant information \autocite{gibson_ecological_1979}. In the context of a generative model that captures our prior assumptions about the world, perception can be conceptualized as \textit{inference} \autocite{von_helmholtz_handbuch_1867}. Probabilistic inference provides a normative perspective on how perception \textit{should} work to make optimal use of limited sensory data. Human vision, in particular, is often conceptualized as an approximation to probabilistic inference on a generative model \autocite{knill_bayesian_2004, friston_theory_2005, yuille_vision_2006}. Given limited neural hardware and compute time, however, it is difficult to implement the normative ideal. The cognitive theories and neural network mechanisms we review here can be understood as heuristic approximations to inference on a generative model.

Cognitive scientists and engineers have begun building models that can maintain internal state and dynamically map the sensory input to internal object representations that have their own persistence and dynamics. Brains and models must decide what qualifies two bits of the visual image to be grouped together as parts of the same object \autocite{treisman_binding_1996, von_der_malsburg_correlation_1981}. Containment within a closed contour and persistence over time of shape, color, and motion are key factors determining how humans segment a scene into objects \autocite{scholl_object_2007, spelke_principles_1990}. These factors are encapsulated by the more general notion of spatiotemporal contiguity, which provides evidence for an underlying physical property: cohesion. But how are the sensory indications of spatiotemporal contiguity combined and their conflicts resolved? How are the object representations untethered from the sensorium, and made to persist when the object disappears behind an occluder? How are they animated jointly by sensory data and generative models of the world? These remain computational mysteries of the human mind and brain.

The focus of this review is on the general computational mechanisms of object-based representations, which are generative and recurrent and complementary to the discriminative feedforward mechanism underlying the initial sweep of activity through the visual hierarchy. We describe these mechanisms in the context of generic rigid bodies. However, these general mechanisms could be replicated in the brain in domain-specific modules that are adapted to the particular properties of behaviorally important objects. Like the feedforward mechanisms that learn the appearance of objects in different domains (such as faces, people, animals, buildings, food, and tools), the object-based mechanisms will additionally adapt to the behavior of the objects, including their ways of moving (e.g., facial expressions), their rigidity (e.g., for rocks and buildings) or articulation (as for bodies and tools), their interactions with other objects (be it according to the laws of classical mechanics or theory of mind), and their behavioral relevance.

We first review behavioral phenomena and cognitive theories of human object representations, and then the current state of neural network modeling. Our goals are to highlight parallels between cognitive concepts and neural network model mechanisms and to discern what characteristics of human object representations are missing in current neural network models. We hope this review will help (1) modelers understand the behavioral literature, (2) behavioral researchers understand the computational literature, and (3) both groups develop tasks that can serve simultaneously as probes of human cognition and as benchmarks for computational models.

\section*{Cognitive Theories}

Cognitive scientists have explored object vision with behavioral experiments, and their concepts and theories summarize the insights gained (Fig. \ref{fig:overview}). \textit{Grouping} of visual features and \textit{amodal completion} yield a rapid initial scene segmentation that transcends the static filters of the feedforward visual hierarchy, but remains tethered to the retinal reference frame. This retinotopic representation forms the basis for selection of a limited set of objects for representation in an object-based reference frame, known as \textit{object-files} or \textit{slots}. At this level, object representations are untethered from the retinal reference frame and may enter central cognition \autocite{duncan_selective_1984, neisser_cognitive_1967, treisman_features_1986} and interaction with other cognitive systems \autocite{baars_cognitive_1993, dehaene_towards_2001}. The cognitive concepts we review here, as of yet, lack full mechanistic specification. However, they help summarize the behavioral phenomena, decomposing the cognitive processes and providing essential stepping stones toward their implementation in neural network models.

\subsection*{Tethered to the retinal reference frame: pixels to proto-objects}

\subsubsection*{Grouping features}

The simplest way to combine evidence over space is using static filter templates. This is the mechanism of models of V1 simple and complex cell responses \autocite{hubel_receptive_1965}. A hierarchy of such filters \autocite{riesenhuber_hierarchical_1999} yields texture statistics at different spatial scales, as employed in convolutional feedforward neural networks \autocite{lecun_backpropagation_1989}. However, there is evidence that the visual system also uses lateral recurrent signal flow to relate collinear edges \autocite{roelfsema_cortical_2006, field_contour_1993, geisler_visual_2008, bosking_orientation_1997}. Dynamic recurrent processing through lateral interactions may provide a more flexible mechanism for grouping features at larger scales. Imagine, for example, the set of all smooth closed contours. The combinatorics of feature configurations forming a smooth closed contour may render representation of this set with a basis of static filters unrealistic. However, the regularity of smooth continuation can be exploited by a model using lateral recurrent connectivity.

Principles of perceptual grouping were first identified by Gestalt researchers \autocite{koffka_principles_1935, rock_legacy_1990, wertheimer_untersuchungen_1923}, who noted that people perceive visual elements as grouped by principles including continuity, proximity, similarity, closure, pr\"{a}gnanz, and common fate. One of these principles, continuity, involves the detection and integration of contour elements \autocite{field_contour_1993}, and the computation of border-ownership for the creation of surface representations \autocite{nakayama_experiencing_1992}. Feedforward \autocite{rosenholtz_intuitive_2009} as well as recurrent operations \autocite{li_neural_1998, yen_extraction_1998} that incrementally group contours by spread of activation \autocite{roelfsema_object-based_1998} have been proposed. Perceptual grouping is influenced by several factors such as binocular disparity \autocite{nakayama_serial_1986}, textures \autocite{julesz_experiments_1975} and temporal coincidence \autocite{alais_visual_1998} and knowledge about object appearances \autocite{vecera_is_1997}.

Local integration processes may give rise to a \textit{mosaic stage} \autocite{sekuler_perception_1992}, in which each connected set of visible parts of an object forms a group. The mosaic stage is similar to Marr’s \autocite{marr_vision:_1982} \textit{full primal sketch}, in which contour integration gives rise to an initial grouping. In Marr's theory, the primal sketch is followed by the 2.5D sketch, which represents the visible portions of objects as surfaces and assigns a depth to each patch of the image. Once surfaces and depth relationships are represented in the 2.5D sketch, the visual system can infer how objects may extend behind occluders. Disjoint mosaic pieces belonging to the same object (disconnected by occlusion) can be grouped together and the occluded parts filled in.

\subsubsection*{Amodal completion}

Visual scenes often contain objects that are partially occluded by other objects. Moreover, objects always occlude their own backsides. We nevertheless perceive them as 3-dimensional wholes. It has been proposed that this subjective experience might result from a process that explicitly fills in the missing parts of an object in our mental representation. The process has been called \textit{amodal completion} \autocite{michotte_nouvelle_1951} because, in contrast to perceptual filling-in (i.e., modal completion) \autocite{komatsu_neural_2006}, it transcends the sensory modality: the occluded part or backside of an object is not \textit{visually} perceived, yet it is part of the percept.

\begin{figure*}[ht]
\centering
\includegraphics{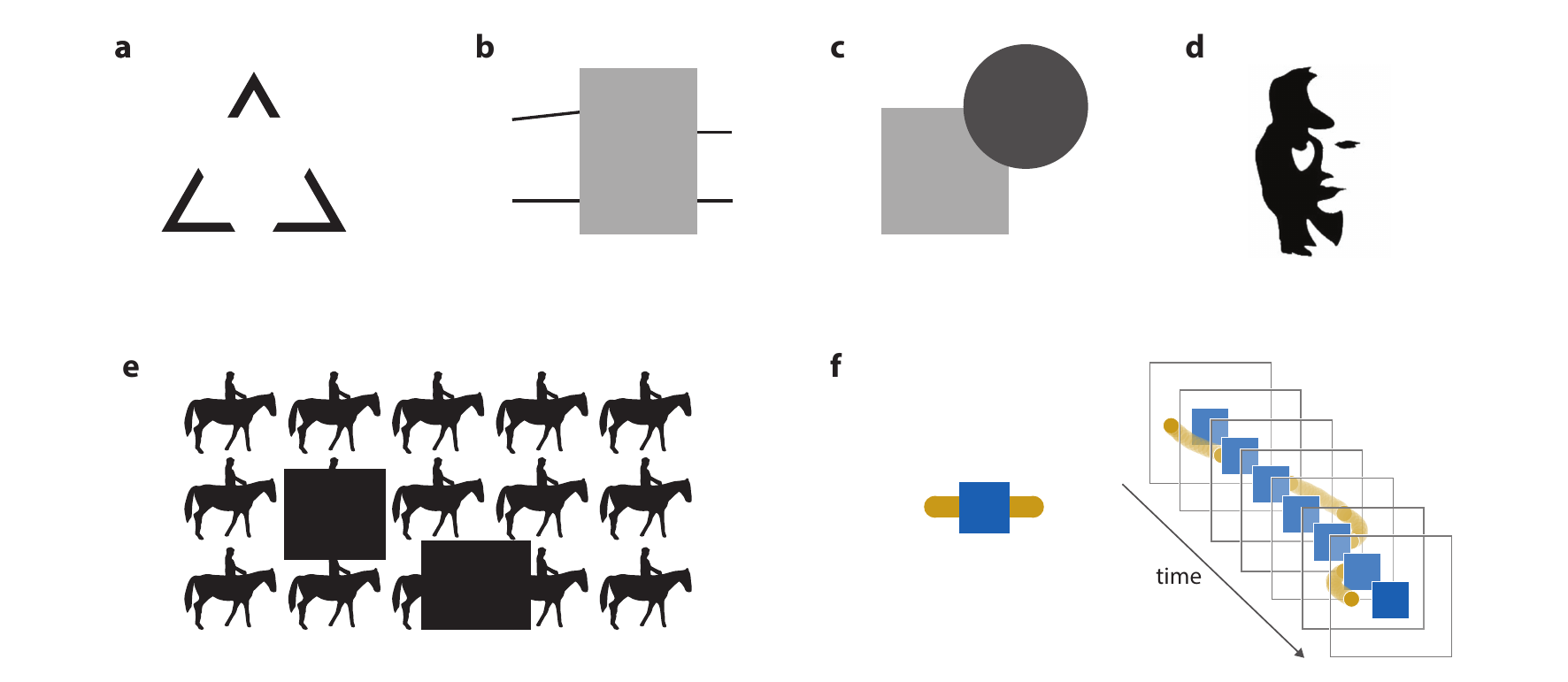}
\caption{\label{fig:amodal_completion} \textbf{Completion phenomena}. (a) There appears to be a solid white triangle occluding the black contours of another triangle. The percept of the occluding white triangle is an example of \textit{modal} completion, because the inferred contours appear as though they were present in the visual modality. The percept of the occluded black triangle is an example of amodal completion, because the missing black contours are perceived to exist, but are not visually perceived. (b) The lower black line segments appear connected behind the gray box. This is an example of amodal completion because the inferred continuation is not perceived as visible in the image. (c) A complete gray square appears to be present. This is an example of amodal completion on the basis of global shape cues. (d) We perceive a face lit from the right. This is an example of perceptual closure \protect{\autocite{mooney_age_1957}}. (e) People may perceive a giraffe-like rider (upper left black box) or an elongated horse (lower right black box) \protect{\autocite{kanizsa_amodale_1970}}. These percepts are inconsistent with both the global repetitive pattern and our prior knowledge about the anatomy of horses and people. Such illusions demonstrate that local cues can override global cues and prior knowledge in the perceptual inference process. (f) On the left, we perceive a single golden object extending behind the blue occluder. This is an example of amodal completion that requires grouping of all the golden bits across space. Perceptual inference can also group bits of visual evidence across space and time simultaneously. On the right, the frames of a movie are shown, where a golden ball oscillates behind a blue occluder. When watching such a movie, we perceive a persistent object whose presence continues across periods of total invisibility. Our visual system groups the golden bits into a "space-time worm". This is an example of spatiotemporal amodal completion.}
\end{figure*}

Beyond the phenomenology of subjective experience, the hypothesis of an amodal completion process suggests testable behavioral predictions. A partially occluded object should elicit priming effects that match those elicited by its complete form, rather than those elicited by its visible fragments (Box \ref{box:tasks}e). This prediction has been confirmed in behavioral experiments \autocite{sekuler_perception_1992}. Similar predictions have been confirmed for discrimination \autocite{shore_shape_1997} and visual search tasks \autocite{he_surfaces_1992, rensink_early_1998}. These studies have also shown that it takes time for amodal completion to emerge, suggesting that it relies on recurrent processing  \autocite{sekuler_perception_1992,shore_shape_1997}.
 
Amodal completion must rely on prior knowledge. It could use general knowledge about the statistics of images (e.g. the knowledge that edges tend to extend smoothly) or about the shape of objects (e.g. the knowledge of the shape of an occluded part of a letter). It could also rely on knowledge gleaned moments earlier from having observed the now occluded parts of the object. There is evidence that amodal completion extends edges behind occluders if a continuous smooth connection exists \autocite{kellman_theory_1991}. Amodal completion is also thought to fill in missing parts of surfaces \autocite{he_surfaces_1992} and volumes \autocite{tse_volume_1999}. Local completion extends and connects object contours mostly linearly according to the Gestalt principle of good continuation (Fig. \ref{fig:amodal_completion}b). Global completion refers to completion that prefers symmetric solutions (e.g., Fig. \ref{fig:amodal_completion}c) \autocite{buffart_coding_1981} likely occurring in higher visual areas such as the lateral occipital complex \autocite{weigelt_separate_2007,thielen_neuroimaging_2019}. More generally, the term \textit{perceptual closure} \autocite{mooney_age_1957, snodgrass_priming_1990} refers to completion based on prior knowledge about the shape or appearance of an object (e.g., Fig. \ref{fig:amodal_completion}d).

Amodal completion may best be construed as an inference process: the visual system’s best guess about the missing part, given the current evidence and prior knowledge. The computational function of making the inferred information explicit might be to support further inferences about the object.

\subsubsection*{Proto-objects}

The initial input segmentation occurs in parallel and pre-attentively across the visual field \autocite{neisser_cognitive_1967, treisman_feature-integration_1980}. These processes are largely independent of conscious cognition, in the sense that our conscious thoughts cannot penetrate and interfere with them \autocite{pylyshyn_is_1999}. For example, consciously thinking that the horse pattern in Fig. \ref{fig:amodal_completion}e should extend regularly behind the occluder does not prevent the visual system from generating the percept of an elongated horse.

These initial segmentations are thought to be tethered to the retinal reference frame. As a consequence, they are subject to change whenever we move our eyes or the world evolves. Moreover, the grouping of features might not yet be definitely established at this early stage. It might be best understood as a set of tentative feature associations than a full parse of the scene into object representations \autocite{wolfe_psychophysical_1999}. Hence, these representations have been termed \textit{proto-objects} \autocite{rensink_dynamic_2000}, to acknowledge their volatile and tentative nature. Transforming a proto-object representation into a stable and spatiotemporally coherent object-based representation will require selection by higher cognitive processes and untethering from the retinal reference frame.

\subsection*{Untethered from the retinal reference frame: object files and pointers}

In order to individuate objects and combine the distributed evidence about them, the visual system has to overcome a fundamental challenge: How to group the spatiotemporally disjoint pieces into a coherent object representation? In the retinal reference frame, the pieces had to be grouped in space. Now the grouping problem extends in space and time. Rather than segmenting retinal space, the system must carve out a ``space-time worm'' \autocite{scholl_object_2007} from the spatiotemporal input (Figure \ref{fig:amodal_completion}f)).

How does the visual system link distinct sensory inputs across occlusions or saccades to a single object-centered representation? In many situations, this \textit{correspondence} problem \autocite{ullman_interpretation_1979} is solved by assessing the spatiotemporal continuity of objects \autocite{hood_spatiotemporal_2009, mitroff_space_2007, scholl_object_2007}. A striking example is the \textit{‘tunnel effect’} \autocite{burke_tunnel_1952}. An object that moves behind an occluder and reappears with different appearance (such as a different color or even category) may still be considered to be the same object by the visual system instead of two different ones \autocite{flombaum_temporal_2006, michotte_les_1964}. A single object is more likely to be perceived if the pre-occlusion stimulus is similar to the post-occlusion stimulus \autocite{hollingworth_object_2009, moore_features_2010}, suggesting a general mechanism that flexibly weighs object feature dimensions to infer correspondence \autocite{papenmeier_tracking_2014}. If correspondence is inferred, we perceive a single object whose appearance combines pre- and post-occlusion sensory signals. The post-occlusion appearance of the object is biased toward the pre-occlusion stimulus \autocite{liberman_serial_2016, fischer_context_2020}. Eye-movement studies \autocite{irwin_memory_1992} similarly suggest that both the locations and appearances of stimuli are used to establish correspondences across saccades \autocite{richard_establishing_2008}.

Correspondence computations support stable internal representations of individuated, untethered object representations that transcend the retinal or spatial reference frame. Different cognitive theories have been proposed that encapsulate empirical findings of how object representations might interact with the retinal bound proto-object representational level \autocite{kahneman_reviewing_1992, pylyshyn_role_1989, rensink_dynamic_2000}. These theories emphasize the importance of space over other features to individuate and keep track of objects. Different objects tend not to occupy the same portion of space simultaneously. The natural domain to uniquely track objects across time therefore is the spatial domain. Feature integration theory suggests that segregation of the input into objects and binding of object features to coherent representations occurs via space \autocite{treisman_feature-integration_1980}. Pylyshyn \autocite{pylyshyn_role_1989} proposed an indexing system that individuates and tracks objects via spatial pointers or indices. While \textit{visual indexes} are pointers to locations they themselves encode no object properties. Hence, Pylyshyn termed his theory FINST for ‘fingers of instantiation’ as indices work like physical fingers: without knowing anything about the tracked (pointed to) object, spatial information such as a location or spatial relations between different fingers can be extracted.

Similarly, Kahneman and colleagues \autocite{kahneman_reviewing_1992} proposed that our visual system \textit{individuates} each object by creating an \textit{object-file} that groups a subset of the proto-objects carved out in the retinal reference frame on the basis of spatiotemporal factors. In contrast to visual indices, object-files are thought to also store information about the properties of the object (e.g., color, shape), thus re-representing and \textit{‘binding’} essential sensory information in a coherent object representation \cite{kahneman_reviewing_1992}. This process is termed \textit{identification} because the feature information defines the identity of each object. Evidence for separate processing of object features bound into a coherent object representation comes from studies in which humans perceive \textit{illusory conjunctions} of features of two different objects \autocite{wolfe_psychophysical_1999} under some conditions, demonstrating the failure of the process. The individuation of an object is thought to precede the identification of its appearance, as famously captured by the observation of Kahneman and colleagues \autocite{kahneman_reviewing_1992} that humans can conceive of something as the same ‘thing’ while its identity remains in flux and might dramatically change over time: "Onlookers in the movie can exclaim: ‘It’s a bird; it’s a plane; it’s Superman!’ without any change of referent for the pronoun" (p. 217).

One of the hallmark features of human cognition is that the number of simultaneously maintained object files is highly limited. These capacity limitations are often phrased in terms of limited attentional resources. Spatiotopic maps may encode the distribution of attention over the visual field. These \textit{spatial attention maps} \autocite{itti_computational_2001} may be the access point of the spatial indexing system in which object-files could be created from saliency peaks via center-surround inhibition. Multiple object-files can then each be tracked by top-down attention in the spatial attention map \autocite{cavanagh_tracking_2005}. A mechanistic explanation for the capacity limitation of the object-file system therefore is \textit{surround inhibition} \autocite{bahcall_attentional_1999} between spatial pointers in these maps \autocite{franconeri_flexible_2013}. 

One influential class of tasks that now has been employed in hundreds of empirical studies is \textit{multiple object tracking} \autocite{pylyshyn_tracking_1988} (Box \ref{box:tasks}h). Humans can track a limited number of objects (perhaps three or four) even through full occlusions \autocite{intriligator_spatial_2001, pylyshyn_tracking_1988, scholl_tracking_1999, yantis_multielement_1992}. Subsequent research found that the tracking limitations can better be described by a flexible resource \autocite{vul_explaining_2009} that is independent across hemifields \autocite{cavanagh_tracking_2005}. For example, if slower object speed reduces spatial crowding, up to eight objects can be tracked \autocite{alvarez_how_2007}.

Selection of an object for tracking entails a processing advantage for all of its elements and for the spatial positions it occupies \autocite{flombaum_attentional_2008, yantis_multielement_1992}. This manifests in faster and more accurate detection of targets that appear on tracked compared to untracked objects. The processing advantage extends across the whole representation and suggests that objects are the fundamental units of attentional selection \autocite{vecera_does_1994}. \textit{‘Object-based attention’} benefits both dynamic and static objects \autocite{chen_object-based_2012, duncan_selective_1984, egly_shifting_1994, houtkamp_gradual_2003, jeurissen_serial_2016}, objects that are only partially visible and completed amodally \autocite{moore_object-based_1998}, and even objects that are completely invisible and retained in memory for a brief duration \autocite{peters_activity_2015, peters_object-based_2020}.

\begin{box1}*

\begin{centering}
\includegraphics[width=0.99\textwidth]{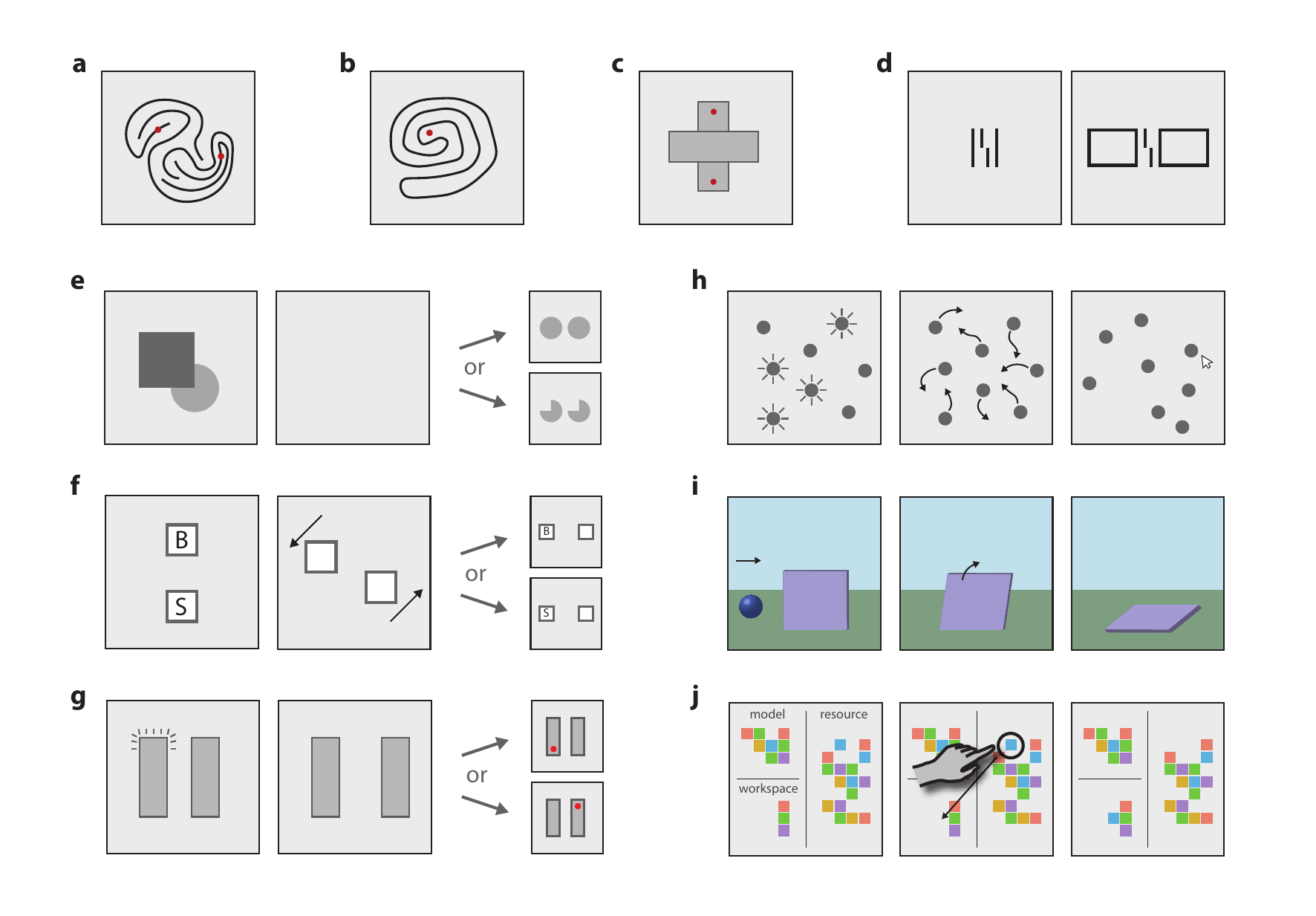}
\end{centering}

{\small
Cognitive scientists have developed a variety of ingenious tasks to probe human untethered object perception with behavioral experiments. Grouping tasks (a-d). Four different tasks for contour integration and grouping. (a) Decide as fast as possible whether two dots lie on the same or different lines \autocite{jolicoeur_curve_1986}. (b) Decide whether the dot lies inside a closed contour \autocite{ullman_visual_1984}. (c) Decide whether both red dots lie on the same object \autocite{pitkow_exact_2010}. (d) Detect the direction of the horizontal offset between the central vertical lines in the presence of flankers. The task is more difficult if the flankers, too, are isolated (crowding, left) and easier if the flankers are part of a coherent object (uncrowding, right) \autocite{sayim_gestalt_2010}. Amodal completion (e). A partially occluded shape (here: a circle) is presented as a prime. Subsequently, participants are presented with two shapes and have to decide whether these are identical \autocite{sekuler_perception_1992}. Responses are faster if these shapes match the percept of the prime (e.g., the circles if the percept was amodally completed). Object-reviewing paradigm (f). In a typical object-reviewing trial \autocite{kahneman_reviewing_1992} two objects containing a letter are presented during the previewing display. In the test display, only one letter is presented and needs to be identified. Reactions are faster if the letter is in the same object as in the previewing frame. Here, the objects also switch positions. Object-based attention (g). In the object-based attention task \autocite{egly_shifting_1994} one end of one object is briefly flashed to attract attention to this position. After a brief delay participants have to react as quickly as possible to a target (red dot). Reactions are faster when the target appears in the same object (top) as the flash than when it appears in the other object (bottom). Multiple object tracking (h). A set of targets is flashed initially and has to be tracked among identical distractors. After the tracking phase, participants have to select the identity of the tracked targets \autocite{pylyshyn_tracking_1988}. Violation of Expectation (i). Violation of expectation to study object permanence and physical reasoning. Here, a solid ball disappears behind a wall that subsequently folds down. Observer’s surprise is measured (e.g., by measuring the looking time) in response to this physically impossible sequence of events (e.g., \autocite{baillargeon_object_1985}). In the block-copy task (j) participants have to reconstruct a model visual pattern in a workspace area using building blocks from the resource area \autocite{ballard_deictic_1997}.
}

\end{box1}

\subsection*{Object permanence, visual working memory, and mental simulation}

Objects can transiently cease to elicit retinal responses, for example when they become occluded and when we shift our gaze. Internal object representations, however, can remain stable even with their links to the input momentarily severed. The knowledge that out of sight is not out of mind has been termed \textit{object permanence} by Piaget \autocite{piaget_construction_1954}. In infants, artificial stimuli that violate object permanence elicit longer looking times, consistent with surprise (violation of expectation, Box \ref{box:tasks}i). The results of such experiments support the idea that a kernel of object permanence may be either innate or established within 3 or 4 months after birth \autocite{baillargeon_object_1987, baillargeon_object_1985, spelke_origins_1992}. However, the ability to represent objects not currently in view likely matures over early development \autocite{wilcox_object_1999, rosander_infants_2004, moore_visual_1978}.

Adults can track objects through full occlusions without noticeable performance decrements \autocite{scholl_tracking_1999}. This suggests a remarkable ability of our visual system to attribute spatiotemporally disjoint sensations to the same coherent object representation. An object representation can better track the sensory signals elicited by its object if it captures the dynamics of its object and predicts its future location and state \autocite{lake_building_2017}. Evidence for \textit{mental simulations} of object dynamics comes from studies of \textit{representational momentum}, which show that people incorrectly estimate the angle of a suddenly disappearing rotating object as slightly advanced along the rotational motion trajectory \autocite{freyd_representational_1984}. The mental simulations seem to be confined to first-order dynamics: Humans appear to use velocity, but not acceleration to simulate objects behind occluders \autocite{benguigui_time--contact_2003, rosenbaum_perception_1975}. From a normative perspective, prediction of the dynamics should be important for an object representation to track its object through longer periods of occlusion so as to find the sensory signals elicited by the object as it re-emerges. However, in most real-world scenarios that humans encounter a coarse, approximate prediction of the dynamics might potentially suffice to successfully track objects. Indeed, psychophysical evidence suggests that human perceptual inferences rely heavily on coarse spatiotemporal heuristics \autocite{franconeri_simple_2012}. 

The fact that object representations can bridge occlusions implies that some information about the object is stored during occlusion. But what is the nature of this internal untethered representation? Another frequent event that momentarily severs the object representations from the sensorium is the saccade, during which input into the visual system is suppressed (\textit{saccadic suppression}, \cite{matin_saccadic_1974}). Asking people to detect changes of visual patterns across saccades reveals that their \textit{transsaccadic memory} is capacity-limited and does not retain detailed spatial information but rather abstract and relational information \autocite{henderson_two_1994, irwin_memory_1992}. 

The limits of human object representations are also evident in multiple-object tracking tasks. When the objects are suddenly occluded, people can recall location and velocity (including direction) information, but not the detailed identifying features of the objects \autocite{bahrami_object_2003, pylyshyn_puzzling_2004}. In particular, shape and color are difficult to consciously recall a moment later \autocite{horowitz_tracking_2007}, although information about them (along with location and velocity) is maintained across occlusions \autocite{hollingworth_object_2009, moore_features_2010}.

These findings suggest that the human visual system does not maintain an object representation that fully specifies all its features. Instead - for the purpose of bridging disruption of the input as caused by saccades or occlusions - only a small subset of the features of an object is maintained. 

A candidate system that can encode and maintain visual information for a limited amount of time during occlusions or saccadic remapping is \textit{visual working memory} \autocite{flombaum_temporal_2006, fougnie_distinct_2006, hollingworth_binding_2010}. This system is severely limited in its capacity. Visual working memory capacity was originally conceptualized as a limited number of \textit{slots} for individual objects (similar to object-files) \autocite{awh_visual_2007, cowan_magical_2001, luck_capacity_1997, miller_magical_1956}. Subsequent research has questioned strong versions of the slots hypothesis. For example, remembered objects don't fail as a unit, rather object features and their bindings to the object can be forgotten independently for the same object \autocite{bays_storage_2011, fougnie_object_2011}. The memory representations may better be characterized as hierarchically structured feature bundles \autocite{brady_review_2011} in which bindings and features can fail independently. The capacity of visual working memory has also been characterized as a limited continuous \textit{resource} that can be divided up among the objects with a different portion allotted to each \autocite{alvarez_capacity_2004, bays_dynamic_2008, wilken_detection_2004}. A related hypothesis is that the object representations interfere with each other within the same substrate \autocite{oberauer_interference_2017, bouchacourt_flexible_2019}. Importantly, the concept of working memory goes beyond mere storage. The ‘working’ part refers to flexible access and control of information for the purpose of higher-order cognitive processes such as visual reasoning \autocite{baddeley_working_1974, cowan_evolving_1988, miyake_models_1999}. 

\section*{Neural network models}

The cognitive theories capture the human behavioral phenomena and provide a blueprint for computational models. However, they fall short of fully specifying the algorithm or how it might be implemented in a neurobiologically plausible way. We now discuss attempts to implement untethered object representations in neural network models. Ever since the inception of the first artificial neuron models \autocite{mcculloch_logical_1943}, researchers have studied how cognitive capacities can arise from the interaction of neurons in a network \autocite{oreilly_computational_2000}. The classic models were designed for small toy problems, raising the question of whether their computational mechanisms scale to real-world vision. Modern computer hardware and software enable us to test these mechanisms in large-scale models that perform real-world visual tasks. A successful example is the deep convolutional mechanism, which was first implemented in the neocognitron \autocite{fukushima_neocognitron:_1980} 40 years ago and which, in the past decade, has enabled deep neural networks to perform image recognition \autocite{krizhevsky_imagenet_2012, lecun_backpropagation_1989}.

\subsection*{Neural network mechanisms and cognitive phenomena}

Multi-layer perceptrons \autocite{rosenblatt_principles_1961, rumelhart_learning_1986-1, ivakhnenko_polynomial_1971} and their convolutional variants \autocite{fukushima_neocognitron:_1980}, including modern deep convolutional neural networks \autocite{lecun_backpropagation_1989}, lack mechanisms for untethered object representation. However, the classic literature also has a rich history of models that implement mechanisms for untethered object representations, such as completion, grouping, object files, and working memory. We first outline some elemental mechanism for associative completion, gating, routing, and grouping and describe how neural networks may represent untethered objects and perform probabilistic inference. We then consider how these elements may interact to implement the cognitive functions of modal and amodal completion, object files and slots, and object permanence.

\begin{box2}*

{\small
he binding problem refers to a set of computational challenges of how different elements can flexibly and rapidly be linked to each other in a network, where connections change only at the slow time scale of learning. Binding has often been studied in the context of vision, where it refers to binding of parts and properties of objects, objects to locations, and objects across time \autocite{treisman_binding_1996}. Binding is not a problem intrinsic to vision but results from the specific implementation of a visual system. For example, when different features of the same object (e.g., color and shape) are preferentially analyzed in separate, specialized regions, they might need to be linked or recombined together subsequently again. Several solutions to the binding problem in neural networks have been proposed \autocite{oreilly_three_2003, greff_binding_2020, von_der_malsburg_correlation_1981, hummel_solution_2004}. For example, specialized neurons could signal the presence of specific feature combinations (i.e., conjunction coding) \autocite{hinton_distributed_1987}. This approach is however limited due to the combinatorial explosion of possible feature combinations and the fact that only previously learned combinations can be represented. Humans however can perceive and act upon arbitrary and previously unseen feature combinations (e.g., “Consider seeing a three-legged camel with wings, or a triangular book with a hole through it”, \autocite{treisman_solutions_1999}, p. 108). Distributed representations of conjunctions that encode feature combinations in a coarse code \autocite{ballard_parallel_1983} or via tensor product coding \autocite{smolensky_tensor_1990}, or dynamic interunits \autocite{feldman_dynamic_1982} could alleviate these downsides. Instead of using feature combination detectors, a network could dynamically adapt its weights to bind features of the same object together \autocite{schmidhuber_learning_1992}. Another binding challenge arises when simultaneously perceiving multiple objects. As a consequence of increasing receptive field sizes, higher-level visual neurons receive input from the full visual field and potentially from multiple objects at the same time. This superposition in neuronal populations is problematic if the information cannot be uniquely attributed to the different objects (i.e., the superposition catastrophe \autocite{von_der_malsburg_am_1986}). How does the brain distinguish between these multiple objects in a distributed representation? One solution may be to sequentially process individual objects \autocite{olshausen_neurobiological_1993, reynolds_role_1999, tsotsos_modeling_1995}. In the brain, such temporal multiplexing of object representations could be implemented in theta rhythmic neural activity \autocite{fries_rhythms_2015}. In addition, this selective processing of individual proto-objects might be necessary to bind constituent features into a structural description of the object \autocite{roelfsema_cortical_2006, treisman_feature-integration_1980, wolfe_psychophysical_1999}. A prominent and highly debated proposal of how the brain solves the binding problem is the idea that binding is expressed via correlated activity of neural assemblies that encode the same object \autocite{gray_stimulus-specific_1989, hummel_dynamic_1992, von_der_malsburg_correlation_1981}. Neurons could operate as coincidence detectors of synchronous incoming spikes of feature detectors that represent parts which should be bound together, temporarily increase synaptic efficacy for these inputs, and decrease sensitivity to asynchronous inputs (but see \autocite{shadlen_synchrony_1999}). The temporal phase at which feature detectors spike then represents a dimension that labels the temporary grouping a neuron belongs to.
}
\end{box2}

\begin{figure*}[ht]
\centering
\includegraphics[width=0.98\textwidth]{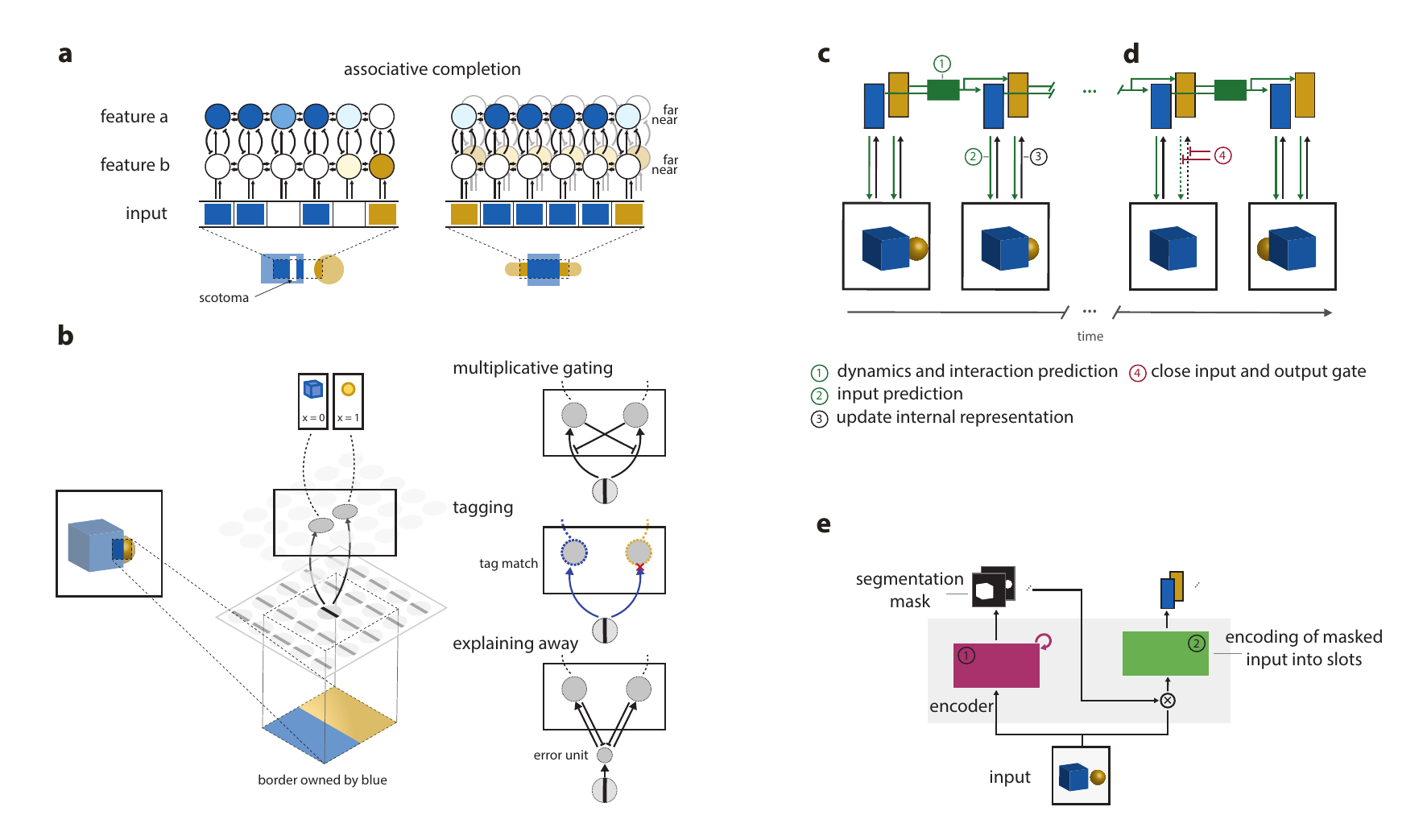}
\caption{\label{fig:network_mechanisms} \textbf{Neural network mechanisms for untethering}. \textbf{(a) Associative completion} can fill in missing information. Here a scotoma is bridged in the representation via lateral connections which perform \textit{modal} completion (left). Associative processes may also contribute to \textit{amodal} completion (right), which additionally requires units for different depth planes (right). \textbf{(b) Local routing mechanisms} enable context-dependent local modulation of the connectivity between units at the time-scale of inference. The network layer detects the presence of an edge, which in this case belongs to the blue object. Gating mechanisms selectively route information to the part of the network that represents the blue object. Three gating mechanisms are illustrated. \textit{Multiplicative gating} suppresses the input to the units not representing the target object. \textit{Tagging} adds a label (e.g., a temporal or phase tag) to the activation (here blue lines indicate a tag corresponding to the blue object), which is used by upstream units to filter their inputs. \textit{Explaining away} subtracts already explained parts from the input \protect{\autocite{rao_predictive_1999}}. \textbf{(c) Predictive processing with structured representations} engages multiple mechanisms. Prediction of dynamics and interactions between objects occurs at the level of object representations (e.g., slots) (1). The prediction at the abstract level of the latent representation may be decoded into lower-level predictions that are closer to the input at the next time step (2) and object representations are updated depending on the prediction error (3). \textbf{(d) Memory gating}. During occlusion, the yellow object is persistent and has to untether its connection with the input (4). \textbf{(e) Global routing via recurrent spatial attention.} Example for separate localization and encoding of objects in a DNN \protect{\autocite{burgess_monet:_2019, eslami_attend_2016}}. A recurrent attention network computes segmentation masks which select portions of the image for routing into separate object slots.}
\end{figure*}

\paragraph*{Associative completion.}
If a neuron or model unit were to implement a feature detector, it would be useful for it to listen to its neighbors for evidence that its feature is present or absent. When two features are correlated in natural visual experience, bidirectional connections with equal weights between the neurons representing the two features can help both neurons detect their features in the presence of noise (Fig. \ref{fig:network_mechanisms}a). Such connectivity could be acquired by Hebbian learning \autocite{hebb_organization_1949}.

The prevalence of smooth contours in natural images renders approximately collinear edge detectors correlated under natural stimulation \autocite{geisler_visual_2008}. There is evidence that V1 neurons selective for collinear edge elements are preferentially connected by excitatory synapses \autocite{bosking_orientation_1997}. The lateral connections may implement a diffusion process that regularizes the representation, shrinking it back toward a prior over natural images or collapsing behaviorally irrelevant variability, so as to ease the extraction of relevant information by downstream regions.

Symmetric lateral connectivity can also implement autoassociative completion of complex learned patterns \autocite{hopfield_neural_1982}. The weight symmetry enables us to understand the dynamics of the network in terms of an energy function. An activity pattern far from all of the learned patterns will have high energy. From such a point in state space, the dynamics will descend the energy landscape until it reaches a fixed-point attractor, a local minimum of the energy function, corresponding to one of the learned patterns \autocite{zemel_localist_2001, iuzzolino_convolutional_2019}. 
Associative completion can more generally be understood as predictive regularization. When the predictions are not just across space (as in the example above), but also across time, they can approximate a Bayes filter, which optimally combines past and present evidence. The connection weights between two units will not be symmetric then, and the dynamics, rather than converging to fixed-point attractors, can model the dynamics of the environment \autocite{deneve_optimal_2007}. Such a mechanism might implement the cognitive phenomenon of representational momentum \autocite{freyd_representational_1984}.

Associative completion processes could be used not just within, but also across levels of the visual hierarchy. In either case, associative completion involves interactions between units that directly adjust what we may think of as the units' representational content. Next we consider a complementary set of mechanisms that operate at a higher level: modulating \textit{interactions} between units, rather than unit activity, so as to gate, route, and group the representational content.

\paragraph*{Gating, routing, and grouping.}

Object representations could be inferred from the input by a set of static filters. However, this approach would require filters for all possible shapes, sizes, and locations of objects and their interactions when one partially occludes another. A more efficient solution with respect to the number of units needed is to use static filters for parts (in particular parts that are frequently encountered) and to dynamically compose the parts to represent a given object. The composition can be implemented by selectively \textit{routing} lower-level part representations to the higher-level representation of the object. Architectural connections in a neural network between units representing parts, then, are \textit{potential connections}, a subset of which is instantiated to represent a specific object. This requires a routing mechanism: a rapid modulation of the connectivity between units at the time-scale of inference \autocite{schmidhuber_learning_1992}. An example of routing is a neural-shifter circuit that dynamically maps retinal input from varying locations into a location-invariant (i.e. object-centered) representation \autocite{olshausen_neurobiological_1993, anderson_shifter_1987, burak_bayesian_2010}.

Routing can be implemented by multiplicative modulation of the input gain to a unit \autocite{salinas_gain_2000, hochreiter_long_1997}. During grouping, units can influence the gain functions of other units that compete to explain the same lower-level input. The unit that wins responsibility for the input may end up closing the gate between the input and the other competing units (Fig. \ref{fig:network_mechanisms}b).

Instead of attenuating the connectivity between units, a neural network might also use explicit tagging of messages. For example, the message that a neural activation conveys (e.g. the presence of a feature) could be tagged with a signal indicating which group it belongs to \autocite{reichert_neuronal_2014}. A receiving unit could then selectively combine information over inputs with the relevant tag (Fig. \ref{fig:network_mechanisms}b). One such mechanism that has been investigated in neuroscience is binding-by-synchrony, in which a temporal tag is provided by the time of firing, and units that fire synchronously are considered as signalling features of the same object \autocite{gray_stimulus-specific_1989, hummel_dynamic_1992, von_der_malsburg_correlation_1981, fries_rhythms_2015}.

Another form of gating is subtractive gating, where input to a unit is canceled by inhibition from a gating unit. For example, predictive coding \autocite{rao_predictive_1999} employs a process of subtractive \textit{explaining away}, where higher-level units explain their lower-level input and subtract their predictions out of the lower-level representation (Fig. \ref{fig:network_mechanisms}b). What remains are the unexplained portions of the lower-level representation, the residual errors, which continue to drive the higher-level units. The resulting recurrent dynamics can implement an \textit{iterative inference} process, in which higher-level units converge to a state where they jointly account for the input. A higher-level unit that explains a part of the input (e.g., an object that clutters or partially occludes another object) will explain away its portion of the image, preventing that portion from interfering with the recognition of the other portions. Predictive coding combines forms of routing and grouping, processing the image in parallel, but successively accounting for more of the objects and their interactions as it progresses from the easy to the hard parts.

\paragraph{Untethered representation of objects.}

We refer to object representations as \textit{untethered} if they are free from immediate control by the sensory stimulus. Untethered representations can combine information over time scales, including recent sensory information (e.g. about the trajectory of an object as it moved behind an occluder) and prior knowledge (e.g. about the behavior of objects of a category). To exploit the objects' relative independence in the world, untethered object representations must \textit{disentangle} the information about different objects \autocite{higgins_towards_2018}. One approach is to dedicate a separate set of units, a \textit{neural slot}, to the representation of each object. Alternatively, multiple objects can be represented in a shared population of units as \textit{distributed representations}. Each unit might have \textit{mixed coding} for different objects, but the information about different objects could still occupy separate linear subspaces. For both slot and mixed representations, the object representations may be distributed across hierarchical levels that jointly encode a scene-parsing tree, \autocite{feldman_what_2003, hoffman_parts_1984, hummel_dynamic_1992} with lower levels encoding detailed features and higher levels more abstract aspects of the object.

\paragraph{Probabilistic inference on a generative model.}

A neural network implementation of probabilistic inference on a generative model must combine probabilistic beliefs \autocite{pouget_probabilistic_2013} about the latent variables (the prior) with the probability of the sensory data given each possible configuration of latents (the likelihood) \autocite{lee_hierarchical_2003, friston_theory_2005, rao_predictive_1999}. The generative model would need to specify the prior over the object-level representation and how to generate an image from that representation. Perception then amounts to inversion of the generative model, inferring the object-level representation from an image. Assuming we are given the generative model, we might train a feedforward neural network to approximate the mapping from data to posterior, using training pairs of images and latents obtained either by drawing latents from the prior and generating images \autocite{dayan_helmholtz_1995} or by using a generic inference algorithm to infer latents from images drawn from some distribution. Speeding up inference by memorizing past inferences is called \textit{amortization} \autocite{stuhlmuller_learning_2013}. A feedforward neural network can memorize frequently needed inferences and generalize to novel inferences to some extent. However, for complex generative models, the stochastic inverse may not lend itself to efficient representation in a feedforward network with a realistic number of units and weights. Fully leveraging the generative model for generalization may require generative model components to be explicitly implemented and dynamically inverted during perceptual inference, which requires recurrent computations \autocite{van_bergen_going_2020}. Challenges with probabilistic inference include the acquisition of the generative model and the amount of computations required for inference. Brains and machines must strike some compromise, combining the statistical efficiency of generative inference with the computational efficiency of discriminative inference. For example, instead of evaluating the likelihood at the level of the image, the inference may evaluate the likelihood at a discriminatively summarized higher level of representation. In addition, short of inference of the full posterior, a network may use a generative model to infer only the most probable latent variable configuration for a specific input, the maximum a posteriori (MAP) estimate \autocite{rao_predictive_1999}. One approach is to seed the inference with a first guess about the objects and their locations computed by a feedforward computation. The initial estimate can then be iteratively refined toward the MAP estimate. At each step, the likelihood can be evaluated by synthesizing a reconstruction of the sensory data using a top-down network that implements the generative model.

\paragraph*{Inferring object properties beyond the visible input.}

The associative completion described above can fill-in missing pieces or otherwise repair a representation corrupted by undesirable variability (including internal and external noise, as well as behaviorally irrelevant variation of the objects). Perhaps surprisingly, elaborating the representation through memory, regularizes the representation, and thus reduces the information about the stimulus. This may be desirable if the information lost is not relevant. If associative completion is to collapse undesirable variability, it should overwrite the sensory representation. This may explain illusory contours and other \textit{modal} completion phenomena \autocite{von_der_heydt_searching_2003} (Fig. \ref{fig:network_mechanisms}a). Associative completion might also contribute to \textit{amodal} completion. For example, the occluded portion of a contour of a simple convex shape could be extrapolated locally using prior assumptions about contour shape (e.g., an assumption of smoothness). Whether associative completion can by itself explain amodal completion phenomena, however, is questionable \autocite{kogo_side_2013}. An associative mechanism for amodal completion would require dedicating a different set of units to the inferred, but invisible features. Separate units for inferred features would enable the system to represent the occluder and the occluded parts of the back object simultaneously in different depth planes. More generally, separate units for inferred features might help a probabilistic inference process avoid confusing inferred features for independent sensory evidence.

Alternatively or in addition to associative completion, amodal completion phenomena may arise through the representation of the object as a whole at a higher level. The same mechanisms \autocite{craft_neural_2007, grossberg_neural_1985, mingolla_neural_1999, zhaoping_border_2005} that group the visible features, by combining priors about object shape with sensory information, might also give rise to the percept of an amodally completed object. Higher-order priors on object shape can be implemented in a hierarchical neural network. For example, a hierarchical neural network based on the neocognitron \autocite{fukushima_neocognitron:_1980} has been shown to infer occluded contours via \textit{feedforward} and \textit{feedback} interactions \autocite{fukushima_neural_2010}.

When we conceptualize the visual system as performing generative inference \autocite{yuille_vision_2006}, amodal completion can be considered an emergent phenomenon resulting from inference about whole objects from partial input. Here, gating and routing mechanisms that instantiate dynamical assignments during hierarchical, iterative inference are particularly important. Lower-level units that respond to the visible parts of a partially occluded object activate units at the next higher level that represent the hypothesis that the object is present. The likelihood of this hypothesis can be evaluated by feedback connections that predict the presence of the full object at the lower, part level \autocite{zhuowen_tu_image_2002}. Such predictions will not match the evidence at the site of occlusion, unless the representation of the occluder explains away the occluded portion \autocite{fukushima_restoring_2005,lucke_occlusive_2009}. 
Alternatively, a feedback-controlled gating mechanism could restrict the evaluation of the likelihood of the presence of the partially occluded object to the unoccluded portion. With either mechanism, the occluder-induced gating prevents the absence of evidence for the object where it is occluded from being misinterpreted as evidence of absence of the object. This is consistent with the fact that occlusions, but not deletions induce amodal completion \autocite{johnson_recognition_2005}.

\paragraph*{Representing and tracking multiple objects}

When multiple objects need to be represented or tracked by object-based representations, an accounting mechanism may be helpful that ensures a one-to-one mapping between slots and objects. Ensuring a one-to-one mapping prevents interference between features of different objects (the superposition problem, Box \ref{box:binding}). This can be implemented by different routing mechanisms. One approach is temporal multiplexing, the separation of different objects in time. Temporal multiplexing can operate at a fine temporal scale, with precise spike synchrony \autocite{gray_stimulus-specific_1989} or a shared oscillatory phase \autocite{reichert_neuronal_2014, fries_rhythms_2015}, indicating that two signals belong to the same object. Alternatively, temporal multiplexing can operate at a coarse temporal scale, for example when covert or overt attention sequentially selects different objects \autocite{koch_shifts_1987, olshausen_neurobiological_1993, tsotsos_modeling_1995, walther_modeling_2006}). As an alternative to temporal multiplexing, a unique frequency \autocite{kazanovich_oscillatory_2006} can be used to tag an object slot and avoid interference with objects represented by other slots. For any of these tagging mechanisms, an inhibitory mechanism between slots can ensure that each slot is assigned a unique tag. In the framework of predictive coding, one-to-one mappings can dynamically emerge through error representations and explaining away. Tracking of objects across time can be achieved by combining the prior prediction of the object's position with the incoming sensory evidence.

\paragraph*{Bridging spatiotemporal gaps.}

As an object moves, it might become occluded by other objects. When it disappears behind an occluder and reappears on the other side later on, the spatiotemporal gap in the stream of visual evidence may be too large for local mechanisms, such as lateral associative filters, to bridge. The gap induced by a full occlusion of the object also severs the established routing between the sensory signals and the object-based representation. How can an object slot reestablish its correspondence to the sensory evidence after such a gap?

An object could be tracked through occlusion via a model-based temporal filter that continuously simulates its hidden state (including its motion and other property transformations) through the period of full occlusion. At the same time, a mechanism is needed that prevents the visual input from the occluder from interfering with the representation of the hidden object. This can be accomplished by a gating mechanism or by recurrent dynamics that separate sensory and mnemonic contents into different linear subspaces of a neural representation \autocite{libby_rotational_2021}. Correspondence with the sensory stream could be reestablished if the object reappears within the margin of error of the simulated position.

A short-term memory mechanism can maintain the hidden object state while the object is occluded. Several mechanisms have been proposed to explain how information is maintained in a network over a limited amount of time \autocite{barak_working_2014, durstewitz_neurocomputational_2000}. The most popular class of model proposes that recurrent dynamics retain information in \textit{attractor states} \autocite{compte_synaptic_2000, wang_synaptic_2001, wimmer_bump_2014, zenke_diverse_2015}. Such mechanisms have been used to model object permanence in infants. The mechanism predicts the disappearance of an object behind an occluder, dynamically maintains the representation of the object while it is invisible, and predicts its re-appearance \autocite{mareschal_computational_1999, munakata_rethinking_1997}. 

Short-term memory is a central requirement not just for object tracking, but for many cognitive tasks. An alternative to active maintenance is \textit{activity-silent storage}, which could be supported by short-term plasticity of connections. The activity representing the object can be restored upon retrieval \autocite{mi_synaptic_2017, mongillo_synaptic_2008}. Recently, both active and activity-silent mechanisms have been shown to dynamically interact in short-term memory depending on task demands \autocite{masse_circuit_2019}.

Beyond information storage, short-term memory also needs to support flexible updating of content, retrieval of a subset of the information for ongoing computations, and selective deletion \autocite{chatham_multiple_2015}. Like object tracking, these operations require a gating mechanism \autocite{frank_interactions_2001, gruber_dopamine_2006, oreilly_biologically_2006} that can rapidly grant access to a stored memory or protect its content from interference (Fig. \ref{fig:network_mechanisms}d). The long short-term memory \autocite{hochreiter_long_1997} and related gating mechanisms have been successfully employed to address this problem.

\subsection*{Modern deep neural networks as models of human object vision}

The neural network mechanisms for untethered object perception described in the previous section were often implemented in small models that could only handle toy tasks. Candidate mechanisms for explaining human vision need to scale to real-world tasks. The breakthroughs with deep convolutional neural networks \autocite{krizhevsky_imagenet_2012, ciresan_multi-column_2012} and the associated hardware and software advances have provided the technological basis for addressing this challenge \autocite{schmidhuber_deep_2015,lecun_deep_2015}. 

Modern deep neural network models are typically constructed by training an architecture on a particular objective using backpropagation. The neural mechanisms emerge from the interplay of the architecture, the optimization objective, the learning rule, and the training data. On the one hand, learning is necessary for a complex model to absorb the knowledge and skills needed for successful performance under real-world conditions. A vision model, for example, needs to learn what things look like. On the other hand, the fact that the neural mechanisms emerge through learning renders a trained model with millions of parameters somewhat mysterious, motivating post-hoc investigations into its mechanism \autocite{zhou_interpreting_2019}. Modelers do exert control over the mechanisms, but at a more abstract level: by designing the architecture, the optimization objective, the learning rule, and the training experiences \autocite{richards_deep_2019}.

It is an open question whether brains can use backpropagation or a related error-driven learning rule \autocite{crick_recent_1989, lillicrap_backpropagation_2020, kording_supervised_2001, guerguiev_towards_2017, scellier_equilibrium_2017, roelfsema_attention-gated_2005}. Whether or not it is biologically plausible, backpropagation can serve as a tool to set the parameters of models meant to capture the computations underlying perceptual performance. When we use it as such, we forgo any claims as to how the interaction of genes, development, and experience produced such solutions in humans. Ultimately, of course, we would also like to understand how a biological visual system incorporates visual experience on the longer timescales of learning and development, and to model this process with a biologically plausible learning algorithm.

Modern deep neural networks scale up many of the known neural network mechanisms. Feedforward convolutional neural networks (CNNs) have been very successful in tasks such as visual object recognition \autocite{krizhevsky_imagenet_2012, szegedy_inception-v4_2017}. The architecture of CNNs \autocite{fukushima_neocognitron:_1980, lecun_backpropagation_1989} is inspired by the primate visual hierarchy. CNNs capture many aspects of cognitive and neuroscientific theories of pre-attentive parallel visual processing. They integrate information over a hierarchy of spatial or spatiotemporal filters, with filter templates replicated across spatial positions. When trained to recognize object categories, their internal representations are similar to those of the human and nonhuman primate ventral visual stream \autocite{khaligh-razavi_deep_2014, guclu_deep_2015, yamins_performance-optimized_2014, kriegeskorte_deep_2015,yamins_using_2016}.

The best computer-vision models for object recognition so far are deep CNNs. However, CNNs lack many of the mechanisms of human object perception. For example, it has been shown that these networks rely more strongly on texture than humans, whose recognition prominently depends on global shape information \autocite{baker_deep_2018, geirhos_imagenet-trained_2019, brendel_approximating_2019}. CNNs see the image in terms of summary statistics that pool local image features, which provides a surprisingly powerful mechanism for discriminating object categories. However, they do not decompose the scene into objects, or objects into their parts, as is required for the model to understand the structure of the scene (AI objective) and to explain human cognitive phenomena, such as amodal completion and object files.

Computer vision must solve many tasks beyond texture-based recognition, such as localization, instance segmentation \autocite{he_mask_2017, pinheiro_learning_2015}, and multiple object tracking (e.g., of pedestrians, sports players, vehicles, or animals) \autocite{luo_multiple_2021}. Like the human visual system, these models must localize, individuate, identify, and keep track of multiple objects. They employ computational strategies broadly similar to those in the cognitive literature. For example, object localization models \autocite{girshick_rich_2014} use region-proposal methods, a strategy similar to the saliency maps of the visual system \autocite{bisley_attention_2010, koch_shifts_1987}, and sequential instance segmentation and recognition of objects \autocite{burgess_monet:_2019, locatello_object-centric_2020} (Fig. \ref{fig:network_mechanisms}e), which resembles the cognitive theory of sequential individuation and identification \autocite{kahneman_reviewing_1992}. Computer vision also uses global shifts of attention as a form of temporal multiplexing to infer multiple objects \autocite{eslami_attend_2016}. Computer-vision systems often combine learned CNN components with hand-crafted higher-level mechanisms like physics engines \autocite{wu_learning_2017}, providing interesting hybrid (cognitive and neural) models that could be tested formally as models of human vision. However, it is also important to pursue more organically integrated RNN models that can maintain representations over time, sequentially attend to different portions of the visual input, and individuate, identify and track multiple objects.

Models more consistent with human object vision can can be developed by introducing constraints at each of Marr's three levels of analysis: \autocite{marr_vision:_1982} the level of biological implementation, the level of representation and algorithm, and the level of the computational objective. We consider these three levels in turn.

\subsubsection*{Constraints from neurobiology}

Deep CNNs provide a coarse abstraction of the feedforward computations performed by the human visual system. However, they do not have lateral and top-down recurrent connections, and therefore lack the ability to maintain representations over time \autocite{van_bergen_going_2020}. RNN models trained on object recognition provide better models of human brain representations and behavior than deep feedforward networks \autocite{spoerer_recurrent_2017, kubilius_brain-like_2019, kietzmann_recurrence_2019, spoerer_recurrent_2020}. Segmentation, identification, and amodal completion of object instances are naturally solved by iterative algorithms that can be implemented in recurrent networks. This may explain why neural networks endowed with recurrence yield better performance in object recognition under challenging conditions such as occlusions \autocite{oreilly_recurrent_2013, wyatte_early_2014, spoerer_recurrent_2017}. Biologically inspired gating of lateral connections has been shown to yield more sample efficient training during tasks like segmentation \autocite{linsley_sample-efficient_2018}. Neurobiology continues to provide rich inspiration for modeling work that will explore the computational benefits of more realistic model units, architectural connectivity, and learning rules.

\subsubsection*{Constraints on representations and algorithms}

The space of possible solutions an RNN may implement for a particular task is large. Object-based representations or generative inference do not automatically emerge through task training. Modelers have therefore endowed their architectures with representational structure thought to reflect aspects of the generative structure of the world. For example, models use neural slots at the latent level for inference in static images and in dynamic tasks \autocite{eslami_attend_2016, engelcke_genesis:_2020, steenkiste_relational_2018, wu_learning_2017, burgess_monet:_2019, greff_multi-object_2019}. Slots are attractive because they are interpretable and provide a strong inductive bias for task-trained models. However, slots may fall short in capturing phenomena such as illusory conjunctions \autocite{wolfe_psychophysical_1999} or the capacity limitations of human cognition \autocite{cowan_magical_2001, miller_magical_1956}, which can manifest in gradual degradation of the fidelity with which objects are represented as the number of objects grows \autocite{alvarez_capacity_2004, bays_dynamic_2008, wilken_detection_2004}. Representing a variable number of objects in a shared neural population resource \autocite{swan_binding_2014, oberauer_interference_2017, schneegans_neural_2017, matthey_probabilistic_2015} combined with binding mechanisms (Box \ref{box:binding}) promises to explain these cognitive phenomena.

Modelers can also constrain the inference algorithm by imposing hierarchical representations \autocite{sabour_dynamic_2017, xu_unsupervised_2019}. Inference in capsule networks \autocite{sabour_dynamic_2017, kosiorek_stacked_2019} is based on the idea that the visual input can be segmented into hierarchical groupings of parts. The recurrent inference process decomposes a scene into a hierarchy of parts \autocite{feldman_what_2003, hoffman_parts_1984, hummel_dynamic_1992}. This is accomplished by a routing mechanism that enhances the connectivity between the lower-level capsule and the corresponding higher-level capsule while attenuating connectivity to competing higher-level capsules thereby implementing ``explaining away''. Humans and feedforward neural network models both struggle to recognize objects in visual clutter, a phenomenon known as \textit{visual crowding} \autocite{pelli_uncrowded_2008} (Box \ref{box:tasks}d). However, human recognition of the central object is undiminished if the visual clutter can be ``explained away'' as part of other objects. This \textit{uncrowding} effect \autocite{sayim_gestalt_2010} has recently been demonstrated for capsule networks \autocite{doerig_capsule_2020}, which separate the clutter from the object by representing each in a different capsule.

Discrete relational structures can be expressed in a graph, where objects and parts are nodes and edges represent relations. \textit{Graph neural networks} provide a general and powerful class of model that can perform computations on a graph using neural network components \autocite{battaglia_relational_2018, scarselli_graph_2009}. A softer way to impose structure is to encourage the emergence of a disentangled representation through a prior on the latent space \autocite{higgins_towards_2018, hsieh_learning_2018}. A key question for current research is how structured representations and computations may be acquired through experience and implemented in biologically plausible neural networks \autocite{whittington_tolman-eichenbaum_2020}. 

\subsubsection*{Constraints on the computational objective}

Recent modeling work has moved beyond supervised training objectives, such as mapping images to labels. Rooted in theories of biological reinforcement learning, \textit{deep reinforcement learning} requires weaker external feedback (just a reward signal), making it more realistic as a model of how an agent might learn through interaction \autocite{sutton_reinforcement_2018, botvinick_reinforcement_2019}. In the absence of any feedback, an agent can use \textit{unsupervised learning}, aiming to capture statistical dependencies in the sensory data. An agent interested in all regularities, not just those that are useful for a specific task, will learn a generative model of the data and can base inferences on the more comprehensive understanding provided by such a model \autocite{von_helmholtz_handbuch_1867, rao_predictive_1999, friston_theory_2005}. To learn all kinds of regularities, an agent may challenge itself with its own games of prediction. In \textit{self-supervised learning}, the model learns to predict portions of the data from other portions across time and space (e.g., the future from the past and vice versa, the left half from the right half and vice versa) \autocite{lecun_power_2018}. The ability to learn without any feedback may be essential for acquisition of knowledge that generalizes to novel tasks.

Self-supervised learning techniques have reinvigorated the construction of complex generative models of images and videos \autocite{kingma_auto-encoding_2013, rezende_stochastic_2014, goodfellow_generative_2014}. Although the "true" generative model of visual data is intractable, these models learn rich compositional structure to meet their training objectives, such as predicting upcoming video frames. Object representations provide a natural way to compress and predict the physical world, rendering compression and prediction promising objectives for unsupervised learning of object representations \autocite{schmidhuber_neural_1991, schmidhuber_deep_2015}. 
Nevertheless, learning object-based representations by self-supervision still appears to require strong structural inductive biases on the generative model \autocite{weis_unmasking_2020}.

Even for a simplified generative model of real-world visual data, inferring the posterior over the latents is intractable. Most deep generative models amortize the inference into a feedforward recognition model. The human brain most likely employs a balance between amortized inference using a feedforward mechanism and iterative generative inference using a recurrent mechanism\autocite{van_bergen_going_2020}. Neural network models with object representations that combine amortized and generative inference \autocite{greff_multi-object_2019, veerapaneni_entity_2020} may more closely capture the inference dynamics of the human visual system. Discovering good latent representations and approximate inference algorithms will require bringing together the perspectives of engineering, neuroscience, and cognitive science. 

\section*{Toward neural network models with untethered object representations}\label{sec4}

The cognitive and modeling literatures present the pieces of the puzzle: the cognitive component functions and potential neural mechanisms. Now we have to put the pieces together and build models of how humans see the world as structured into objects under natural conditions. This will require a new scale of collaboration among cognitive scientists and engineers.

Two key components of this endeavor are \textit{tasks} and \textit{benchmarks}. A task is a computer-simulated environment that an agent (a human, other animal, or computational model) interacts with through an interface of perceptions and actions. Computer-administered tasks give us control of all aspects of the interaction. We can design the task world: its perceptual appearance, the set of actions available, and the objectives and rewards.

Tasks lend direction to cognitive science and AI by posing well-defined challenges that provide stepping stones and enable us to measure cognitive performance. In cognitive science, a task carves out what behaviors are under investigation. In AI, a task defines the engineering challenge. If cognitive science and engineering are to provide useful constraints for each other, it will be essential that they engage a shared set of tasks. Tasks should be designed and implemented for use in both human behavioral experiments and neural network modeling \autocite{watters_modular_2021, leibo_psychlab:_2018}. To allow for training and testing of models, stimuli and task scenarios should be procedurally generated to enable production of an infinite number of new experiences.

Tasks form the basis for behavioral benchmarks for models: model evaluation functions that define progress and enable us to select and improve models. We now discuss how new tasks and benchmarks shared among cognitive scientists and engineers can drive progress.

\subsection*{Tasks to train and test untethered object perception} 

\begin{figure*}[ht!]
\centering
\includegraphics[width=0.98\textwidth]{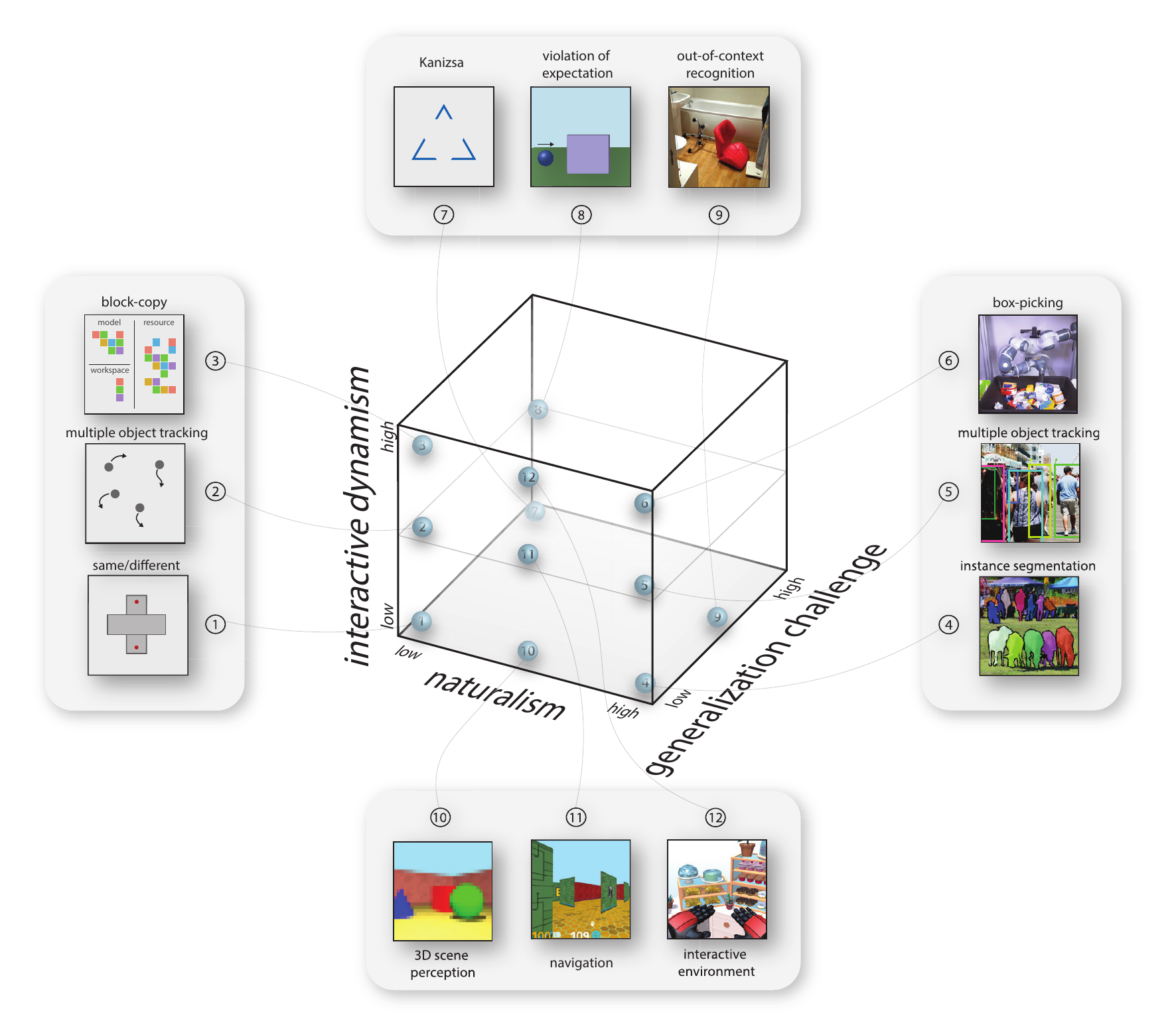}
\caption{\label{fig:new_tasks} \textbf{Space of tasks for untethered object perception}. (a) Three particularly important dimensions of the space of tasks are: naturalism, interactive dynamism, and generalization challenge. \textbf{Naturalism} (horizontal axis): Tasks can be rendered naturalistically or abstracted to their essence. Tasks used in cognitive science (1-3) and machine learning (4-6) tend to concentrate at opposing poles of the naturalism axis. Computer-simulated environments and virtual reality enable us to bridge this gap (10 \& 11: dm-lab \protect{\autocite{beattie_deepmind_2016}}, 12: A2I-THOR \protect{\autocite{kolve_ai2-thor_2017}}). \textbf{Interactive dynamism} (vertical axis): This axis summarizes the degree of dynamism of the stimuli (e.g., movie versus static image) and responses (e.g., motion trajectory versus button press) and the degree of interactivity (i.e., the rate and balance of sensory and motor information flow). Static stimuli as in grouping (1) and segmentation (4) tasks, dynamic stimuli as in multiple object tracking (2, 5), interactive tasks as in the block-copy tasks (3) or box-picking tasks (6, a robot arm has to pick objects from a box with objects). \textbf{Generalization challenge} (depth axis): Tasks can be loosely ordered by the degree to which stimuli are representative of situations encountered during training, be it evolution and learning for the human visual system or the training set used to optimize a neural network model. Tasks that confront the system with untypical (i.e., out-of-training-distribution) situations (7-9, 9: Objectnet \protect{\autocite{barbu_objectnet_2019}}) have high generalization demands and can help reveal the inductive biases of the visual system \protect{\autocite{golan_controversial_2020}}.}
\end{figure*}

Cognitive scientists and engineers tend to design tasks by different criteria, resulting in little overlap in the tasks used. Engineers have focused on tasks that are relevant to real-world applications, often engaging complex natural stimuli and dynamics \autocite{deng_imagenet_2009, geiger_are_2012, sullivan_saycam_2021}. Modeling performance under natural conditions is the ultimate goal. However, complex models are slow to train and difficult to understand. Engineers, thus, should also engage simplified tasks that focus on particular computational challenges. Cognitive scientists often strive to carve cognition at its joints, guided by assumptions about the mind. This has classically led to tasks stripped down to the essential elements required to expose some cognitive component. Simple controlled tasks promise to isolate the primitives of cognitive function \autocite{daw_model-based_2011, pylyshyn_tracking_1988, wilken_detection_2004, green_signal_1966}, rendering behavior directly interpretable in terms of cognitive theory (Box \ref{box:tasks}). However, we must also engage complex and naturalistic tasks to understand how the primitives interact and scale to real-world cognition. Although behavior in complex tasks is harder to interpret per se, it can be used to adjudicate among explicit computational models. Neural networks models, thus, relax the constraint for our tasks to isolate cognitive primitives, liberating us to explore more complex naturalistic task. Even if our tasks do not carve cognition at its joints, they can usefully focus our investigation on a subset of cognitive phenomena whose computational mechanisms are within our reach of understanding.

Cognitive scientists and engineers, then, can benefit from co-opting each other's criteria for a good task. As the former are looking to engage cognition under natural conditions and the latter seek to discover the computational components missing from current AI models, both fields should engage the whole spectrum of tasks, from simple toy tasks to natural dynamic tasks. This strengthens the motivation to collaborate across disciplines on a shared set of tasks. 

Cognitive tasks such as segmentation, visual search, multiple object tracking, physics prediction, or goal-oriented manipulation are good starting points because they focus on plausible cognitive primitives. The world in each of these tasks is a scene composed of persistent objects that can occlude each other and may obey some approximation to Newtonian physics. We here propose to push tasks toward greater complexity along three particularly important axes: naturalism, interactive dynamism, and generalization challenge (Fig. \ref{fig:new_tasks}).

\paragraph{Naturalism.} Naturalism refers to the degree to which the simulated task world resembles the real world. While abstract stimuli are useful for adjudicating among simple models \autocite{rust_praise_2005}, the ultimate goal is to explain perception under natural conditions \autocite{wu_complete_2006}. A synthesis of these two complementary approaches is provided by methods that optimize stimuli to adjudicate among complex models \autocite{golan_controversial_2020}, yielding synthetic stimuli that reflect the natural image statistics the models have learned. For object-based vision, similarly, tasks should achieve various degrees of naturalism while enabling us to adjudicate among models that implement alternative computational theories. We can develop these tasks toward greater naturalism by replacing abstract shapes with photos or 3D models of objects. Incorporating different object categories into these tasks enables us to study the domain specialization of the mechanisms of object perception. For example, tracking of humans and inanimate objects may rely on separate replications of these mechanisms (independent slots) that bring in particular prior knowledge about humans, animals, and inanimate objects.

\paragraph{Interactive dynamism.} Object representations support continuous interaction with a dynamic world (Fig. \ref{fig:overview}). Perception operates at multiple time scales, supporting higher cognitive functions including memory, prediction, and planning. We therefore need tasks that probe performance in dynamic and interactive settings. Cognitive science originally investigated untethered object perception with tasks where a predefined set of static stimuli presented on separate trials elicited a button-press response (e.g., \autocite{ullman_visual_1984, jolicoeur_curve_1986, egly_shifting_1994}, Fig. \ref{box:tasks}). However, more dynamic tasks such as multiple-object tracking \autocite{kahneman_reviewing_1992, pylyshyn_tracking_1988} and interactive tasks such as reproducing an arrangement of blocks (Fig. \ref{box:tasks}, \autocite{ballard_deictic_1997}) have also been developed. In a non-interactive tasks, the initial state is controlled by the experimenter in each of a sequence of trials, rendering behavioral responses easier to analyze and more directly interpretable. When our theories have been expressed in computational models, however, we can also use interactive dynamic tasks to adjudicate among theories. In fact, interactive dynamic tasks will often have a higher bit rate of recorded behavior, promising greater constraints on theory, in addition to enabling us to understand how agents engage dynamic, interactive environments. Task can be pushed from simple toy tasks towards greater interactive dynamism by giving the objects dynamic trajectories and recording responses such as mouse-pointer or eye movements continuously. 

\paragraph{Generalization challenge.} Novel experiences require generalization and are often particularly revealing of the computational mechanism and inductive bias employed by a perceptual system. By probing a model with parameters of the task-generative world that differ from the training distribution, we can generate generalization tests that reveal a model's inductive bias \autocite{geirhos_generalisation_2018, kansky_schema_2017}. To probe untethered object representations, we can present humans and models with novel objects (e.g., procedurally generated 3D models) or with known objects in novel poses or contexts \autocite{barbu_objectnet_2019, weis_unmasking_2020} and study whether task performance generalizes. Tracked objects may change their appearance and shape across time \autocite{blaser_tracking_2000}, which may be hard for models that track by appearance, but easy for humans who primarily track objects based on spatiotemporal properties \autocite{burke_tunnel_1952, hood_spatiotemporal_2009, mitroff_space_2007, scholl_object_2007}. We may also use Gestalt stimuli that elicit grouping in humans (e.g., point light displays of biological motion \autocite{johansson_visual_1973}). We may push our notion of generalization even further to scenarios where there may be no objectively correct response. For example, there is no objectively correct inference to perceive either one or two distinct objects during the Tunnel effect \autocite{burke_tunnel_1952}). However, humans perceive a single object when the spatiotemporal dynamics are consistent with the motion of a single object, revealing the implicit prior assumption that objects are more likely to change than to vanish and appear. Cognitive scientists have probed human perceptual inductive biases with hand-designed stimuli and controlled tasks. These form the basis for generative models of stimuli and tasks that will enable us to comprehensively test and compare generalization behavior in humans and machines.

\subsection*{Benchmarks to evaluate models}

Tasks form the basis for defining behavioral benchmarks for models. A benchmark is an evaluation function that enables us to select and improve models, and to define progress. Engineering has relied on overall task-performance benchmarks \autocite{deng_imagenet_2009}. However, a benchmark can also be defined to measure how close a model comes to emulating human patterns of success and failure across different stimuli and contexts \autocite{schrimpf_brain-score_2018, judd_benchmark_2012, kummerer_understanding_2017, ma_neural_2020, peterson_human_2019, leibo_psychlab:_2018, golan_controversial_2020}. For dynamic interactive tasks, each behavioral episode of a human or model generates a unique trajectory of stimuli and responses. A major challenge is to define useful summary statistics that enable comparisons among humans and models.

Summary statistics can be based on patterns of responses or performance in a task, such as multiple-object tracking, physical reasoning \autocite{bakhtin_phyre_2019}, physical scene understanding \autocite{yi_clevrer_2020, riochet_intphys_2018, baradel_cophy_2020, girdhar_cater_2020}, goal-directed manipulation of objects \autocite{bakhtin_phyre_2019, allen_rapid_2020}, or navigation \autocite{beyret_animal-ai_2019}. A qualitative description such as "performs mental physics simulation" or "can do object tracking" only provides a coarse characterization of a cognitive process. Benchmarks should be based on summary statistics that provide rich quantitative signatures of behavior (e.g., tracking performance as a function of the number of objects to be tracked and other context variables), revealing how humans differ from models \autocite{allen_rapid_2020, riochet_intphys_2018}. Psychophysics and cognitive psychology have developed an arsenal of ingenious methods to probe object perception in humans (Box \ref{box:tasks}), providing much inspiration for the development of benchmarks measuring the behavioral similarity between models and humans \autocite{geirhos_generalisation_2018, ma_neural_2020}. 

\section*{Conclusion}

Perceiving the world around us in terms of objects provides a powerful inductive bias that links perception to symbolic cognition, and action, and forms the basis of our causal understanding of the physical world. Object percepts form through a constructive process of interaction among stages of representation. Deep neural network models have begun to capture components of the process by which object percepts emerge, including grouping, segmentation, and tracking. They do not yet capture the interplay between these components and the powerful abstract inductive biases of human vision. A common set of tasks and benchmarks will help cognitive scientists and engineers join forces. For our models to achieve human-level performance, we will need to be interested not only in the successes, but also in the detailed patterns of failure that characterize human vision.

\section*{Acknowledgements}
B.P. has received funding from the European Union’s Horizon2020 research and innovation programme under the Marie Skłodowska-Curie grant agreement No 841578.

\section*{Competing interests}
The authors declare no competing interests.

\printbibliography

\end{document}